\documentclass [12pt]{article}
\usepackage {amssymb}

\begin {document}

\title {On the information completeness \\
of quantum tomograms}

\author{Grigori G. Amosov\\
Department of Higher Mathematics, MIPT
\\
Dolgoprudny 141700, Russia
\and
Stefano Mancini\\
Dipartimento di Fisica, Universit\`a di Camerino, \\
Camerino 62032 , Italy
\and
Vladimir I. Man'ko\\
PN Lebedev Physical Institute, \\
Moscow 119991, Russia}

\date{\today}

\maketitle

\begin {abstract}
We address the problem of information completeness of quantum measuremets in connection to quantum state tomography and with particular concern to quantum symplectic tomography. We put forward some non-trivial situations where informationally incomplete set of tomograms allows as well the state reconstruction provided to have some a priori information on the state or its dynamics.
We then introduce a measure of information completeness and apply it to
symplectic quantum tomograms.
\end {abstract}

\section {Introduction}

The problem of how to achieve a kind of measurement that is
``complete" in the sense that it can be used to infer information
on all possible (also exclusive) observables dates back to
Ref.\cite{Prug}. Obviously enough, no set of sharp observables
can be informationally complete, while a set of (partially)
non-commuting unsharp observables can be informationally complete
\cite{Busch}. Obviously enough a set of informationally complete
quantum measurements also allows the quantum state
reconstruction. Hence, the connection with quantum state
tomography. Actually, the problem of determining minimal sets of
informationally complete obsevables is equivalent to the group
theoretical problem of finding quantum tomographic schemes (still
unsolved in its generality \cite{Cass}). Following to the approach
introduced in \cite {Vent1, Vent2} probability representations
forming the quantum tomogram can be represented as scalar
products of a state with some elements of abstract Hilbert space.
The information completeness can be checked from this
point-of-view also.

Here we shall consider a quantum system on infinite dimensional Hilbert space $\mathcal{H}=L^2(\mathbb{R})$.
Any measurement is characterized by a positive
operator valued measure (POVM) $\hat M$ infering a positive
operator $\hat M(\Omega )$ to each Borel set $\Omega \subset
{\mathbb R}$. Then, the result of measurement on a state
$\hat \rho $ by means of POVM $\hat M$ is a probability
distribution defined by $P(\Omega )=Tr(\hat \rho \hat
M(\Omega ))$. The question that arises is: how much POVMs $\hat M^{\epsilon
}$, labeled by the parameter $\epsilon$, we should know to
reconstruct the state $\hat \rho $ from a set of probability
distributions $P^{\epsilon }(\Omega )=Tr(\hat \rho \hat
M^{\epsilon }(\Omega ))$?

As we shall show a number of POVMs $\hat
M^{\epsilon }$ which is informatially complete, in the sense that
it allows the reconstruction of the state $\hat \rho $ from
$P^{\epsilon }$, depends on $\hat \rho $. If we use for $\hat
M^{\epsilon }$ the orthogonal resolutions of the identity for the
linear combinations of the position and momentum operators $\hat{X}=\mu
\hat x+\nu \hat p$ [$\epsilon =(\mu ,\nu )$ with $\mu,\nu\in\mathbb{R}$], then the
densities $\omega (X,\mu ,\nu )$ of probabilty distributions
$P^{\epsilon }=P^{\mu ,\nu}$ are said symplectic quantum
tomograms \cite{Mancini}. By fixing a positive number $r$ and putting $\mu =r\cos \theta
,\ \nu =r\sin \theta $, it is known that the set of all rotated position distributions $\omega (X,\theta
)=\omega (X,r\cos\theta , r\sin\theta )$ is informationally
complete, because there is a one-to-one correspondance between
the set of tomograms $\{\omega (X,\theta),\ \theta \in
[0,\pi]\}$ and the quantum state of the system\cite {Vogel}.
This implies the need of an infinite number tomograms to get information completeness. However, in practice these are never available. Hence, it would be important to identify situations
where an incomplete set of symplectic tomograms allows as well the state reconstruction.
It would be also important to quantify the information completeness of a set of tomograms.

Here we shall put forward non-trivial situations
where an incomplete set of symplectic tomograms allows as well the state reconstruction, provided to have some a priori information on the state or its dynamics.
Moreover, we shall introduce a measure of information completeness and we apply it to
symplectic quantum tomograms.

\section {State reconstruction by incomplete knowledge
of symplectic tomograms}

In the symplectic quantum tomography one takes a two-parameter set
$\epsilon =(\mu ,\nu)$ of the POVM's $\hat M^{\epsilon}$
constructed by means of the orthogonal resolutions of the identity
for the observables $\hat{X}=\mu \hat x+\nu \hat p$ \cite{Mancini}.

Let us define a two-parameter set of unitary transforms
${\mathcal F}_{\mu ,\nu}$ in the space $\mathcal{H}=L^{2}({\mathbb R})$ by
the formula
$$
({\mathcal F}_{\mu ,\nu}\psi )(x)=\frac {1}{\sqrt {2\pi|\nu
|}}\int \limits _{\mathbb R} e^{-i\frac {xy}{\nu}+i\frac {\mu
y^{2}}{2\nu}}\psi (y)dy,\ \nu \neq 0.
$$
If $\psi \in \mathcal{H}$ is a wave function in the coordinate
representation, then symplectic quantum tomograms $\omega(X,\mu ,\nu)$
corresponding to the pure state $|\psi\rangle\langle\psi |$
can be written as
$$
\omega (x,\mu ,\nu)=||{\mathcal F}_{\mu ,\nu}\psi ||^{2}
$$

Ideally information completeness is achieved with an infinite number of the above tomograms.
Nevertheless, in some cases a finite number of the above tomograms
(Incomplete knowledge of tomograms) might suffice for quantum state reconstruction,
provided to have some a priori information about the state or its dynamics.

Below we present some non-trivial situations.

\subsection {Finite number of nodes}

Let us consider a particle moving in a one-dimensional potential
$V(x)$.  In the following we shall
claim to know that

\begin{enumerate}

\item[(a)]
the position probability distribution $\omega
(t,X,1,0)\equiv \omega (X,1,0)$ has $M$ nodes at the initial time
$t$.

\end{enumerate}

 We shall call a state $|\psi\rangle\langle\psi|$ compatible with $\omega (X,1,0)$ if
$|\langle X|\psi (t)\rangle|^{2}=\omega (t,X,1,0)$ and $|\langle X|\psi (t+\delta
t)\rangle|^{2}=\omega (t+\Delta t,X,1,0)$ infinitesimally, i.e. for
$\Delta t\to 0$.

We denote by ${\mathcal A}$ and $-\infty
=x_{0}<x_{1}<x_{2}<\dots <x_{M}<x_{M+1}=+\infty$ the set of
possible states compatible with the distribution $\omega (X,1,0)$
and the nodes of $\omega (X,1,0)$, respectively. It was shown in
\cite{Weigert} that the evolution equation for $\omega (X,1,0)$
results in the inclusion $|\psi\rangle\langle\psi |,|\xi \rangle\langle\xi |\in
{\mathcal A}$ iff there exist phases $\phi _{k}\in
[0,2\pi ],\ 1\le k\le M,$ such that
\begin {equation}\label {Con}
\langle X|\xi \rangle=e^{i\phi _{k}}\langle X|\psi \rangle,\qquad X\in [x_{k-1},x_{k}],
\end {equation}
$1\le k\le M+1$.

Now we suppose to know one more tomogram $\omega (X,\mu
,\nu),\ \nu \neq 0$. Then, we introduce the notation
$$
\psi _{j,\mu,\nu}={\mathcal F}_{\mu ,\nu}(\chi
_{[x_{j-1},x_{j}]}\psi),
$$
where $\chi_{[x_{j-1},x_{j}]}=1$ in the interval $[x_{j-1},x_{j}]$ and zero otherwise.
Taking into acount
equality (\ref{Con}), we can write
\begin {equation}\label {phase}
\sum \limits _{j\neq k}e^{i(\phi _{k}-\phi _{j})}\psi _{k,\mu
,\nu }(X)\psi _{j,\mu ,\nu}^{*}(X)=\omega (X,\mu ,\nu)-\sum
\limits _{j=1}^{M+1}|\psi _{j,\mu ,\nu}(X)|^{2}.
\end {equation}

This means that by knowing $n$ tomograms $\omega (X,\mu _{s},\nu
_{s}),\ 1\le s\le n$, we can write a system of equations of the
form (\ref{phase}) for unknown phases $\phi _{k},\ 1\le k\le
M+1$. It is shown in \cite{Weigert} that the matrix $(\psi _{j,\mu
\nu}(X)\psi _{k,\mu ,\nu}(X)^{*})_{j,k=1}^{M+1}$ is invertible if
$n=M$, then there exists a unique solution to the system
(\ref{phase}) and the information given by distributions $\omega
(X,1,0),\ \omega (X,\mu _{s},\nu _{s}),\ 1\le s\le n,$ is
complete.

\subsection {Finite number of different phases}

Let us suppose there exist a fragmentation $-\infty
=x_{-m}<x_{-n+1}<...<0=x_{0}<x_{1}<x_{2}<\dots
<x_{n-1}<x_{n}=+\infty $ and a collection of numbers $0\le \phi
_{j}<2\pi ,\ 1\le j\le n$ such that
$$
\langle X|\psi \rangle=\sum \limits _{j=-n+1}^{n}e^{i\phi _{j}}\chi
_{[x_{j-1},x_{j}]}\psi _{j}(X),
$$
where $\psi _{j}(X)\ge 0,\ X\in {\rm supp}\;\psi
_{j}=[x_{j-1},x_{j}],\ 1\le j\le n+1$. In the following we shall
claim to know

\begin{enumerate}

\item[(a)]
the fragmentation $\{x_{j},\ -m\le j\le n\}$;

\item[(b)]
the square of the wave function $|\psi (X)|^{2}=\omega (X,1,0)$.

\end{enumerate}

Our purpose is to study how can we reconstruct the state by means
of the partial knowledge about tomograms $\omega (X,\mu
,\nu)=\omega (X,r\cos \theta ,r\sin \theta)$. We shall show that
if a number of different phases equals $m+n$, we only need $m+n$
additional angles $\theta _{j}$ for which we know $\omega
(X,r_{j}\cos \theta _{j},r_{j}\sin \theta _{j})$ to reconstruct
the state.

First notice that the conditions (a) and (b) allow us to
reconstruct the functions $\psi _{j}$ as follows
$$
\psi _{j}(X)=\chi _{[x_{j-1},x_{j}]}\sqrt {\omega (X,1,0)}.
$$
Then, let us define a family of functions $\psi _{j,\mu ,\nu}$ by
the formula
$$
\psi _{j,\mu ,\nu}={\mathcal F}_{\mu ,\nu}\psi _{j},\qquad \nu \neq 0.
$$
Since ${\mathcal F}_{\mu ,\nu}$ is a unitary transformation, we
get
$$
\langle\psi _{j,\mu ,\nu}|\psi _{k,\mu ,\nu}\rangle=\langle\psi _{j}|\psi_{k}\rangle=\delta _{jk}.
$$
Moreover,
\begin {equation}\label {trans}
\omega (X,\mu ,\nu)=\sum \limits _{j=-m+1}^{n}|\psi _{j,\mu
,\nu}(X)|^{2}+2\sum \limits _{j\neq k}{\rm Re}(e^{i(\phi _{j}-\phi
_{k})}\psi _{j,\mu ,\nu}(X)\psi _{k,\mu ,\nu}^{*}(X)).
\end {equation}
Denoting $a_{jk\mu \nu}(X)={\rm Re}(\psi _{j,\mu ,\nu}\psi _{k,\mu
,\nu}^{*})$ and  $b_{jk\mu \nu}(X)={\rm Im}(\psi _{j,\mu ,\nu}\psi _{k,\mu
,\nu}^{*})$, we can rewrite Equation (\ref {trans}) as
\begin {eqnarray}\label {trans2}
&&\sum \limits _{j\neq k}\left[a_{jk\mu \nu}(X)\cos (\phi _{j}-\phi
_{k})-b_{jk\mu \nu}(X)\sin (\phi _{j}-\phi _{k})\right]\nonumber\\
&&=\frac {1}{2}\left[\omega (X,\mu ,\nu)-\sum \limits _{j=-m+1}^{n}|\psi
_{j,\mu ,\nu}(X)|^{2}\right].
\end {eqnarray}
The system of functions $\{f_{jk}(X)=\psi _{j,\mu ,\nu}(X)\psi
_{k,\mu ,\nu}(X)^{*},\ j\neq k\}$ can be linearly dependent. Thus, quite generally
it would not be possible to solve Eq.(\ref{trans2}) with respect
to unknown phases. Note however that one can solve this
equation if the time evolution of the system obeys some additional
conditions (see e.g. Ref.\cite{Leonhardt} and the previous section).

Suppose that $|x_{j}-x_{j-1}|=\delta={\rm const}$ is sufficiently
small and $x_{j}=\delta j$, then
$$
\psi _{j}(X)=c_{j}+o(\delta ),\ \delta \to 0,\ x_{j-1}\le X\le
x_{j},\ 1\le j\le n,
$$
where $c_{j}\ge 0$. In such a case we get
$$
\psi _{j,0,\nu}(X)\approx \frac {2\sqrt \nu c_{j}}{\sqrt {2\pi
}X}e^{-i\frac {X\delta j}{\nu }}e^{i\frac {\delta }{2}}\sin \frac
{\delta }{2}
$$
and, for $\mu \neq 0$,
$$
\psi _{j,\mu,\nu}(X)\approx \frac {2\sqrt \nu c_{j}}{\sqrt
{2\pi}X}e^{i\frac {\mu}{2\nu}(X-\delta j)^{2}}e^{-i\frac {\mu
}{2\nu }X^{2}}e^{i\frac {\delta }{2}}\sin \frac {\delta }{2}.
$$
Thus, the functions $f_{jk}=\psi _{j,\mu ,\nu}\psi
_{k,\mu,\nu}^{*}$ are linearly independent for different $j-k$.
Moreover,
$$
\psi _{j,\mu ,\nu}\psi _{k,\mu ,\nu}^{*}=e^{i\frac
{\mu}{2\nu}\delta ^{2}(j^{2}-k^{2})}\psi _{j,0,\nu}\psi
_{k,0,\nu}^{*}.
$$
Notice that $j^{2}-k^{2}=(j-k)(j+k)$ and the system $j-k=r,\
j+k=s$ has a unique solution for the fixed pair $(r,s)$. It
follows that we can pick up $m+n$ angles $\theta _{j}\in [0,2\pi
)$ and write out $m+n$ equations of the form (\ref{trans2}) for
$\mu =r_{j}\cos \theta _{j},\ \nu =r_{j}\sin \theta _{j}$ such
that solving this system we finally obtain the unknown phase
differences $\phi _{j}-\phi _{k}$.

\subsection {Free moving particle}

The Schroedinger equation describing the motion of a free
particle is defined as follows
\begin {equation}\label{ur}
i\psi _{t}=\frac {\hat p^{2}}{2}\psi .
\end {equation}
The solution to (\ref {ur}) is given by the Fresnel integral
\begin{equation}\label{resh}
\psi (X,t)=\frac {1}{\sqrt {2\pi it}}\int \limits _{\mathbb
R}exp\left (i\frac {(X-Y)^{2}}{2t}\right )\psi (0,Y)dY.
\end{equation}
The approach to this situation based upon the Fresnel tomography
was introduced in \cite {Fresnel1, Fresnel2, Fresnel3}. Let us
compare (\ref{resh}) with the Fresnel tomogram $\omega
_{F}(X,\nu)$ (see Formula (8) in \cite {Fresnel1}):
$$
\omega _{F}(X,\nu)=\left |\frac {1}{\sqrt {2\pi i\nu}}\int \limits
_{\mathbb R}exp\left (i\frac {(X-Y)^{2}}{2\nu}\right )\psi
(0,Y)dY\right |^{2}.
$$
It follows that if we know the symplectic quantum tomogram in the
coordinate representation $\omega (t,X,1,0)=|\psi (X,t)|^{2}$ for
all moments of time $t$, then we can reconstruct the Fresnel
tomogram
$$
\omega _{F}(X,\nu)=\omega (\nu ,X,1,0).
$$
It follows that the symplectic quantum tomogram in the initial
moment of time $t=0$ can be derived from the square of wave
function in the coordinate representation $|\psi (X,t)|^{2}$
known for all moments of time by the formula (compare with
Formula (9) in \cite {Fresnel1}):
$$
\omega (0,X,\mu ,\nu)=\frac {1}{|\mu|}\left |\psi \left (\frac
{X}{\mu},\frac {\nu}{\mu}\right )\right |^{2}.
$$
Thus, knowing the dynamics of the symplectic quantum tomogram
only in the coordinate representation we can reconstruct the full
tomogram in the initial moment of time.

\subsection {Parametric driven oscillator}

The dynamical problem of a parametric driven oscillator with frequency
$\omega (t)$, force $f(t)$ depending on time and
Hamiltonian
\begin{equation}\label{Ham}
\hat H(t)=\frac {\hat p^{2}}{2}+\frac {\omega ^{2}(t)\hat
x^{2}}{2}-f(t)\hat x.
\end{equation}
was solved by the method of
time-dependent integrals of motion in Ref.\cite{Manko}.

In the following we shall claim to know

\begin{enumerate}

\item[(a)]
$\omega (t)$, $f(t)$ and that the evolution takes place according to the Hamiltonian (\ref{Ham}).

\end{enumerate}

We denote by ${\mathcal M}(t,X,\mu ,\nu)$ a one-parameter family of
distribution functions associated with the dynamics of quantum
tomograms $\omega _{\hat \rho (t)}(X,\mu ,\nu)$ driven by the
Hamiltonian (\ref {Ham}) such that
$i\hat \rho _{t}=[\hat H ,\hat \rho]$.

Let $\varepsilon (t)$, $\delta (t)$ be functions satisfying the
equations
$$
\ddot \varepsilon (t)+\omega ^{2}(t)\varepsilon (t)=0,
$$
$$
\varepsilon (0)=1,\ \dot \varepsilon (0)=i,
$$
$$
\dot \delta (t)=-\frac {i}{\sqrt 2}\varepsilon (t)f(t),\ \delta
(0)=0.
$$
Then, the dynamics of the distribution functions ${\mathcal
M}(t,X,\mu ,\nu)$ is given by the following formula \cite{Amosov}:
$$
{\mathcal M}(t,X,\mu ,\nu )=
$$
\begin{equation}\label {Evol}
{\mathcal M}(0,X+\sqrt 2 Re((\mu \varepsilon +\nu \dot
\varepsilon)\delta ^{*}),\mu Re(\varepsilon )+\nu Re(\dot
\varepsilon)),\mu Im (\varepsilon) +\nu Im (\dot \varepsilon)).
\end{equation}

This means that if we know the evolution ${\mathcal M}(t,X,\mu ,\nu)$
only for the values $\mu =1$ and $\nu =0$, then we can
get all tomograms $\omega (X,\mu ,\nu)$ at the
initial moment $t=0$ by means of the formula (\ref{Evol}) as
follows
$$
\omega (X,Re(\varepsilon),Im(\varepsilon))=\frac {\partial
}{\partial X}({\mathcal M}(t,X-\sqrt 2 Re(\varepsilon \delta
^{*}),1,0)).
$$

\section {A measure of information completeness}

In many situations
no apriori information is available about the state, hence
it would be helpful to have a measure of information completeness
to use with quantum tomograms.
To define such a measure we
first need to define a measure on convex sets of states.
To this end we exploit the informational measure introduced
and investigated in Ref.\cite{shirokov}.

\subsection {The general case}

Given a statistical ensembele $\{\pi _{j},\hat \rho _{j}\}$
consisting of a probability distribution $(\pi _{j})$ on a set of
states $(\hat \rho _{j})$, one can consider the Holevo $\chi$ quantity
$$
\chi (\{\pi _{j},\hat \rho _{j}\})=S(\sum \limits _{j}\pi
_{j}\hat \rho _{j})-\sum \limits _{j}\pi _{j}S(\hat \rho _{j}),
$$
where $S(\hat \rho)=-Tr(\hat \rho \log \hat \rho)$ is the von
Neumann entropy. Let ${\mathcal A}$ be a convex set of states with
finite von Neumann entropy. Then, the informational measure
of ${\mathcal A}$ is defined by the formula \cite{shirokov}:
$$
\overline C({\mathcal A})=\sup \limits _{\{\pi _{j},\hat \rho
_{j}\}}\chi (\{\pi _{j},\hat \rho _{j}\}),
$$
where the supremum is taken over all probability distributions
$(\pi _{j})$ on subsets of states $\hat \rho _{j}\in {\mathcal
A}$. Notice that $\overline C({\mathcal A})$ is monotonic with
respect to $\mathcal A$, $\overline C({\mathcal A})<+\infty $ iff
$\mathcal A$ is relatively compact and $\overline C({\mathcal
A})=0$ iff the set $\mathcal A$ consists of a single pure state
(\cite{shirokov}, Theorem 2).

Let us suppose that $\hat M^{\epsilon }$ is a set of POVMs labeled by
the parameter $\epsilon $ and that the measurements of the
unknown state $\hat \rho$ by means of $\hat M^{\epsilon}$ result
in the set of probability distributions $P^{\epsilon}$. Then, we consider
the set ${\mathcal A}$ consisting of the states $\hat \sigma $
with the property
\begin {equation}\label {main}
Tr(\hat \sigma \hat M^{\epsilon }(\Omega ))=P^{\epsilon}(\Omega )
\end {equation}
for all parameters $\epsilon $ and Borel sets $\Omega \subset
{\mathbb R}$.

Now, the quantity $\overline C({\mathcal A})$ can be
considered as a measure of information completeness for the set $\hat
M^{\epsilon }$. In particular, if $\mathcal A$ consists of only a
single state, then $\overline C({\mathcal A})=0$ and the set $\hat
M^{\epsilon }$ is informationally complete.

\subsection {Application to symplectic quantum tomograms}

Let us suppose to know the position probability distribution
$\omega (X,1,0)$.  We denote by ${\mathcal A}$ the set of states
generated by all wave functions resulting in the probability
distribution $\omega (X,1,0)$. One can see that $|\psi\rangle\langle\psi|$, $|\phi\rangle\langle \phi|\in{\mathcal A}$ iff they are connected by the
formula
\begin {equation}\label {uni}
\langle X|\psi \rangle=e^{i\xi(X)}\langle X|\phi \rangle,\qquad X\in \rm{supp}\;\omega (X,1,0)
\end {equation}
for some measurable function $\xi(X)$.

Moreover, we suppose to know $N$ additional distributions $\omega
(X,\mu _{n},\nu _{n}),\ \nu _{n} \neq 0,\ 1\le n\le N$. Let
${\mathcal A}$ be the set of states generated by all wave
functions compatible with the distributions $\omega (X,1,0),\
\omega (X,\mu _{n},\nu _{n}),\ 1\le n\le N$.  In this case, we
have $N$ additional relations for $|\psi \rangle\langle\psi |\in{\mathcal A}$
of the following form,
\begin {equation}\label {Sub}
|{\cal F}_{\mu _{n},\nu _{n}}(\psi )(X)|^{2}=\omega (X,\mu
_{n},\nu _{n}),
\end {equation}
$1\le n\le N$. These relations would decrease the
set ${\mathcal A}$ leading in some limit cases to $\overline C(\mathcal
A)=0$.

\subsubsection*{Example: Gaussian states}

A generic zero mean Gaussian state can be described by the characteristic
function \cite{Hol2}
\begin {equation}\label {charact}
\Phi(x,y)=\exp\left[-\frac {1}{2}(\sigma _{xx}x^{2}+2\sigma _{xp}xy+\sigma
_{pp}y^{2})\right],
\end {equation}
where $x,y\in {\mathbb R}$ and the covariances $\sigma _{xx}$, $\sigma _{pp}$, $\sigma _{xp}$ satisfy the Schroedinger-Robertson
uncertainty relation
$$
\sigma _{xx}\sigma _{pp}-\sigma _{xp}^{2}\ge \frac {1}{4}.
$$
The symplectic quantum tomograms corresponding to the
characteristic function (\ref{charact}) are given by
$$
\omega (X,\mu ,\nu)=\frac {1}{\sqrt {2\pi }(\sigma _{xx}\mu
^{2}+2\sigma _{xp}\mu \nu +\sigma _{pp}\nu^{2})^{1/2}}\exp\left
(-\frac {X^{2}}{2(\sigma _{xx}\mu ^{2}+2\sigma _{xp}\mu \nu
+\sigma _{pp}\nu ^{2})}\right ).
$$

Suppose to only know the tomogram
\begin {equation}\label {A}
\omega (X,1,0)=\frac {1}{\sqrt {2\pi \sigma _{xx}}}\exp\left
(-\frac {X^{2}}{2\sigma _{xx}}\right ).
\end {equation}
From it we can retrieve the covariance $\sigma _{xx}$.
Let us calculate the measure $\overline C({\mathcal A})$ for the
set $\mathcal A$ consisting of the Gaussian states compatible
with the distribution (\ref{A}), i.e. with covariance $\sigma _{xx}$. This quantity equals to the
maximum von Neuman entropy overall states in $\mathcal A$ \cite{shirokov}.

In passing we note that the von Neumann entropy of a Gaussian state results \cite {Hol}
\begin {equation}\label {entr}
S(\hat \rho)=g\left(\sqrt {\sigma _{xx}\sigma _{pp}-\sigma
_{xp}^{2}}-\frac {1}{2}\right).
\end {equation}
with
$$
g(x)=(x+1)\log (x+1)-x\log x.
$$

Then, because the condition (\ref{A}) does not restrict the value of
$\sigma _{pp}$ we obtain for our case $\overline C({\mathcal A})=+\infty$.

Now suppose that besides the tomogram (\ref{A}), we also know the tomogram
\begin {equation}\label {B}
\omega (X,0,1)=\frac {1}{\sqrt {2\pi \sigma _{pp}}}\exp\left
(-\frac {X^{2}}{2\sigma _{pp}}\right ).
\end {equation}
From it we can retrieve the covariance $\sigma _{pp}$.
Then, taking into account (\ref {A}),(\ref {B}) and (\ref {entr}) we get
$$
\overline C({\mathcal A})=g\left(\sqrt {\sigma _{xx}\sigma
_{pp}}-\frac {1}{2}\right).
$$

Finally, if we know any other tomogram for additional
parameters $(\mu ,\nu)\neq (1,0)$ or $(0,1)$, it will allow us to
retrieve the covariance $\sigma _{xp}$. Since we supposed a
priori that the set $\mathcal A$ is generated by pure states, we
obtain that our Gaussian state is pure, i.e. $\sigma
_{xp}^{2}=\sigma _{xx}\sigma _{pp}-\frac {1}{4}$, so that
$\overline C({\mathcal A})=0$.

\section {Conclusion}

We have addressed the problem of informational completeness of
quantum measurements in connection to quantum state tomography
and with particular concern to quantum symplectic tomography. We
have put forward some relevant cases where the state
reconstruction is possible by incomplete knowledge of symplectic
quantum tomograms. We have then introduced a measure of
information completeness and we have applied it to symplectic
quantum tomograms. This work sheds further light on the subject
of quantum state characterization which is becoming relevant for
many purposes, e.g. quantum information processing.

\section*{Acknowledgments} The authors are grateful to M.E. Shirokov and S. Weigert for
useful discussions. Grigori Amosov is grateful to Stefano Mancini
for a kind hospitality during his staying at the University of
Camerino. The work of G.A. is partially supported by the INTAS
grant Ref. Nr. 06-1000014-6077 and by the Russian Foundation for
Basic Research under Project No.~07-02-00598.


\begin {thebibliography}{99}

\bibitem{Amosov} Amosov G.G. and Man'ko V.I., Physics Letters A \textbf{318}
(2003) p.287.

\bibitem{Busch} Busch P. and Lahti P.J.,  Ann. der Phys. \textbf{47} (1990)
p.369.

\bibitem{Cass} Cassinelli G., D'Ariano G.M., De Vito E.
and Levrero A., J. Math. Phys. \textbf{41} (2000) p.7940.

\bibitem{Fresnel1} De Nikola S., Fedele, R., Man'ko M.A., Man'ko
V.I. Theor. Math. Phys. \textbf{144} (2005) p.1206.

\bibitem{Fresnel2} De Nikola S., Fedele, R., Man'ko M.A., Man'ko
V.I. J. Russ. Laser Res.\textbf{25} (2004) p.1.

\bibitem{Fresnel3} De Nikola S., Fedele, R., Man'ko M.A., Man'ko
V.I. European J. Phys. B \textbf{36} (2003) p.385.


\bibitem{Hol} Holevo A.S. and Werner R.F., Physical Review A \textbf{63} (2001)
p.032312.

\bibitem{Hol2} Holevo A.S., \emph{Probabilistic and statistical aspects
of quantum theory}, North-Holland Publ. Comp., (1982).

\bibitem{Leonhardt} Leonhardt U. and Raymer M.G., Physical Review
Letters \textbf{76} (1996) p.1985.

\bibitem{Manko} Malkin I.A. and Man'ko V.I., Physics Letters A \textbf{32}
(1970) p.243.

\bibitem{Vent1} Man'ko V.I., Marmo G., Simoni A., Sudarshan
E.C.G., Ventriglia F. Phys. Lett. A \textbf {351} (2006) p.1.

\bibitem{Vent2} Manko V.I., Marmo G., Simoni A., Ventriglia F.
Open Syst. Inf. Dyn. \textbf{13} (2006) p.239.


\bibitem{Mancini}
Mancini S., Man'ko V.I. and Tombesi P., Quant. Semiclass. Opt. \textbf{7} (1995) p.615.


\bibitem{Prug} Prugovecki E., Int. J. Theor. Phys. \textbf{16} (1977) p.321.

\bibitem{shirokov} Shirokov M.E., quant-ph/0510073.

\bibitem{Vogel} Vogel K. and Risken H., Phys. Rev. A \textbf{40} (1989) p.2847.

\bibitem {Weigert} Weigert S., Physical Review A \textbf{53} (1996) p.2078.

\end {thebibliography}
\end {document}